%% file: incoherent_prl.tex
\RequirePackage{lineno}
\documentclass[twocolumn,nofootinbib,prl,superscriptaddress,secnumarabic,amssymb,nobibnotes,aps,floatfix]{revtex4}
\usepackage{graphicx}
\usepackage{hyperref}
\usepackage{amsmath}
\usepackage{dcolumn}
\usepackage{bm}
\newcommand*\patchAmsMathEnvironmentForLineno[1]{%
\expandafter\let\csname old#1\expandafter\endcsname\csname #1\endcsname
\expandafter\let\csname oldend#1\expandafter\endcsname\csname end#1\endcsname
\renewenvironment{#1}%
{\linenomath\csname old#1\endcsname}%
{\csname oldend#1\endcsname\endlinenomath}}%
\newcommand*\patchBothAmsMathEnvironmentsForLineno[1]{%
\patchAmsMathEnvironmentForLineno{#1}%
\patchAmsMathEnvironmentForLineno{#1*}}%
\AtBeginDocument{%
\patchBothAmsMathEnvironmentsForLineno{equation}%
\patchBothAmsMathEnvironmentsForLineno{align}%
\patchBothAmsMathEnvironmentsForLineno{flalign}%
\patchBothAmsMathEnvironmentsForLineno{alignat}%
\patchBothAmsMathEnvironmentsForLineno{gather}%
\patchBothAmsMathEnvironmentsForLineno{multline}%
}

\begin{document}

\title{Exploring the Structure of the Bound Proton with Deeply Virtual Compton Scattering}

\input author_list.tex

\date{\today}
\begin{abstract}
In the past two decades, deeply virtual Compton scattering of electrons has 
   been successfully used to advance our knowledge of the partonic structure of 
   the free proton and investigate correlations between the transverse position 
   and the longitudinal momentum of quarks inside the nucleon. Meanwhile, the 
   structure of bound nucleons in nuclei has been studied in inclusive 
   deep-inelastic lepton scattering experiments off nuclear targets, showing a 
   significant difference in longitudinal momentum distribution of quarks 
   inside the bound nucleon, known as the EMC effect. In this work, we report the first beam spin asymmetry (BSA) 
   measurement of exclusive deeply virtual Compton scattering (DVCS) off a 
   proton bound in $^4$He.  The data used here were accumulated using a $6$ GeV 
   longitudinally polarized electron beam incident on
a pressurized $^4$He gaseous target placed within the CLAS spectrometer in Hall-B at the Thomas
Jefferson National Accelerator Facility. The azimuthal angle ($\phi$) dependence of the BSA was
studied in a wide range of virtual photon and scattered proton kinematics. The 
   $Q^2$, $x_B$,
and t dependencies of the BSA on the bound proton are compared with those on the free proton.
In the whole kinematical region of our measurements, the BSA on the bound proton is smaller by
20\% to 40\%, indicating possible medium modification of its partonic structure.
\end{abstract}

\maketitle 

Electromagnetic probes have played a major role in advancing our knowledge 
about the structure of the nucleon. While lepton-nucleon elastic scattering 
measurements have 
taught us about the spatial charge and magnetization distributions 
\cite{Hofstadter:1955ae,Perdrisat:2006hj}, deep-inelastic scattering 
experiments have uncovered the partonic structure of the nucleon and 
the longitudinal momentum distributions of the constituent partons, i.e., 
quarks and gluons \cite{pdg}.  With nuclear targets, deeply inelastic lepton 
scattering measurements have revealed that the distribution of quarks in a 
nucleus is not a simple convolution of their distributions within nucleons, an 
observation known as the ``EMC effect''\cite{EMC_first} (for reviews on the 
topic, see  \cite{Arneodo:1992wf,Geesaman:1995yd,Norton:2003cb,Hen:2016kwk}).

A wealth of information on the structure of hadrons lies in the correlations 
between the momentum and spatial degrees of freedom of the partons. These 
correlations can be revealed through deeply virtual Compton scattering (DVCS), 
i.e., the hard exclusive lepto-production of a real photon, which provides 
access to a three-dimensional (3-D) imaging of partons within the generalized 
parton distributions (GPDs) framework 
\cite{Mueller:1998fv,Ji:1996ek,Ji:1996nm,Radyushkin:1996nd,Radyushkin:1997ki}.   
The measurement of free proton DVCS has been the focus of a worldwide effort 
\cite{Stepanyan:2001sm,Airapetian:2001yk,Airapetian:2006zr,Chekanov:2003ya,Aktas:2005ty,Chen:2006na,Defurne:2015kxq,Girod:2007aa,Mazouz:2007aa,Gavalian:2009,Seder:2015,Pisano:2015,Jo:2015ema}
involving several accelerator facilities such as Jefferson Lab, DESY and  
CERN. These measurements now enable the extractions of GPDs 
and a 3-D tomography of the free proton \cite{Guidal:2013rya,Dupre:2016mai}.  
New measurements of DVCS from the $^{4}$He nucleus are a critical step towards 
providing a similar 3-D picture of the quark structure of the 
nucleus~\cite{Dupre:2015jha}. In the nuclear case, however, two channels are 
available, the coherent channel where the scattering is off the entire nucleus, 
which is left intact in the final state \cite{Hattawy:2017woc}, and the 
incoherent channel where the DVCS occurs on a nucleon, which is ejected from 
the nucleus. The latter is the focus of this letter and provides a unique 
access to the modification of the partonic structure of the bound nucleons 
\cite{simonetta_2,Guzey:2006xi,Guzey:2008fe}.  The $^{4}$He nucleus is an ideal 
experimental target for this measurement as it is characterized by a strong 
binding energy, a relatively high nuclear core density, and a large EMC 
effect~\cite{JSeely}. Moreover, it remains simple enough that precise 
calculation of its structure can be performed, making this nucleus the perfect 
target for our investigation of the medium modifications of the nucleon's 
partonic structure. The previous measurements of DVCS off nuclei, and in 
particular off $^4$He, performed at HERMES \cite{Airapetian:2009cga} yielded 
results with both "coherent enriched" and "incoherent enriched" event samples, 
hence not fully exclusive, but significant enough to be compared with our 
results below. 

\begin{figure}[tb]
\includegraphics[width=6.5cm]{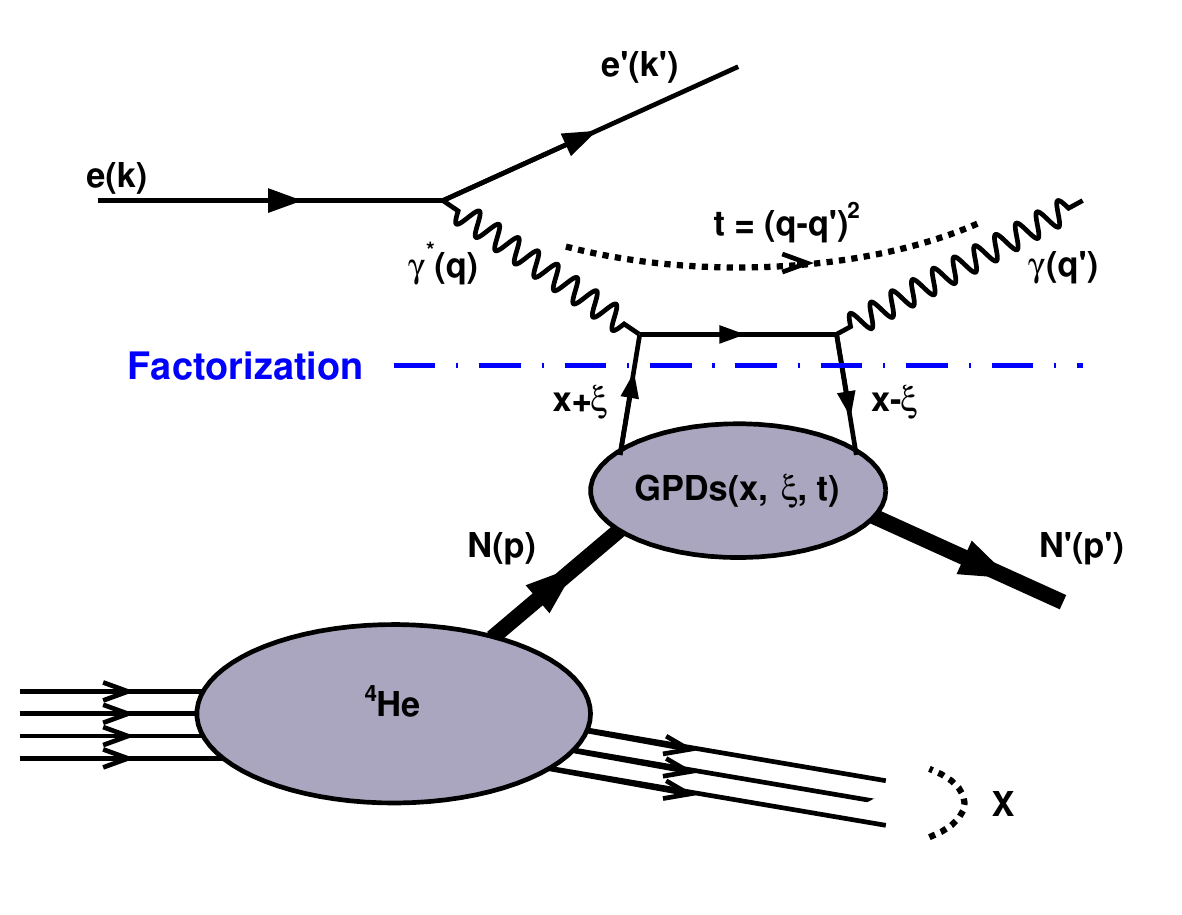}
\caption{Representation of the leading-order, twist-2, handbag diagram of the 
   incoherent DVCS process off $^4$He, where the four-vectors of the electrons, 
   photons, and protons are denoted by $k/k^\prime$, $q/q^\prime$, and 
   $p/p^\prime$, respectively. $x+\xi$ is the nucleon longitudinal momentum 
   fraction carried by the struck quark, -2$\xi$ is the longitudinal momentum 
   fraction of the momentum transfer $\Delta$ ($= q - q^\prime$), and 
   $t$~($=\Delta^2$) is the squared momentum transfer between the initial and
   the final state nucleon.}
\label{fig:diags}
\end{figure}

In this Letter, we present the first exclusive measurement of the beam-spin 
asymmetry (BSA) in deeply virtual electroproduction of a real photon off a 
bound proton in $^{4}$He. Fig.~\ref{fig:diags} illustrates the leading-twist 
handbag diagram for the DVCS process. In the Bjorken regime, i.e. large virtual 
photon four-momentum squared ($Q^{2}=-q^2=-(k-k^\prime)^2$), and at small 
invariant momentum transfer ($t=(q-q^\prime)^2$), the DVCS scattering process 
can be factorized, leaving the non-perturbative structure of the nucleon to be 
parameterized in terms of four chirally even GPDs: $H$, $E$, $\widetilde{H}$, 
and $\widetilde{E}$, representing the four helicity-spin combinations of the 
quark-nucleon states \cite{Freund_Collins,Ji_Osborne}. Experimentally, we 
measure the squared sum of the Bethe-Heitler (BH) and the DVCS amplitudes. The 
BH process, where the real photon is emitted by the incident or the scattered 
electron rather than the nucleon, dominates the cross section at our 
kinematics. The BSA arises from the interference of these two terms and is 
directly sensitive to the DVCS amplitude that contains the information on the 
GPDs. Using a longitudinally polarized electron beam (L) and an unpolarized 
target (U), the BSA is defined as:
\begin{equation}
  A_{LU} = \frac{d^{5}\sigma^{+} - d^{5}\sigma^{-} }
                {d^{5}\sigma^{+} + d^{5}\sigma^{-}},
    \label{BSA_equation}
  \end{equation}
where $d^{5}\sigma^{+}$($d^{5}\sigma^{-}$) is the virtual photoproduction 
differential cross section for a positive (negative) beam helicity. 

Following the cross section decomposition provided in \cite{Belitsky:2001ns}, 
the different components can be expressed in terms of Fourier coefficients 
associated with $\phi$-harmonics, where $\phi$ is the angle between the 
leptonic and the hadronic planes of the reaction. At leading-twist, the BSA can 
be parameterized as: 
\begin{equation}
   A_{LU}(\phi) = \frac{a_{0}\sin(\phi)}{1+a_{1}\cos(\phi)+a_{2}\cos(2\phi)},
   \label{eq:alu-simp}
\end{equation}
where the parameters $a_{0,1,2}$ are combinations of the aforementioned Fourier 
coefficients. The $\sin(\phi)$ harmonic is dominant in $A_{LU}$ and is 
proportional to the following combination of Compton form factors (CFF) 
$\mathcal{H}$, $\mathcal{E}$, and $\tilde{\mathcal{H}}$ as  
\cite{Guidal:2013rya}
\begin{equation}
   a_{0} \propto \operatorname{Im}( F_1 \mathcal{H}- \frac{t}{4M^2} F_2 
   \mathcal{E}+ \frac{x_B}{2}(F_1+F_2)\tilde{\mathcal{H}}),
   \label{a0_cff}
\end{equation}
where $F_1$ and $F_2$ are the Dirac and Pauli form factors, respectively, and 
$x_B$ the Bjorken scaling variable. The real and the imaginary parts of the CFF 
$\mathcal{H}$ relate to the GPD $H$ as  \begin{align}
   \Re(&\mathcal{H}) = \mathcal{P} \int_{0}^{1}dx[H(x,\xi,t)-H(-x,\xi,t)] \, 
   C^{+}(x,\xi), \\
   \Im(&\mathcal{H}) = - \pi [H(\xi,\xi,t)-H(-\xi,\xi,t)],
\end{align}
with $\mathcal{P}$ the Cauchy principal value integral and $C^{+}$ a 
coefficient function defined as $(1/(x-\xi) + 1/(x+\xi))$, where $\xi$ is the 
skewing factor and can be related to $x_B$ by $\xi\approx {{x_B}\over{2-x_B}}$.  
Similar expressions apply for the GPDs $E$, $\widetilde{H}$, and 
$\widetilde{E}$ \cite{Guidal:2013rya}. At the forward limit, $\xi\to 0$ and $t \to 0$, the GPD $H$ reduces to quark, anti-quark PDFs, and its zeroth moment in $x$ represents the elastic Dirac form-factor F$_1$.


The experiment (E08-024~\cite{Hafidi:2008pr}) took place in Hall-B of Jefferson 
Lab using the nearly 100\% duty factor, longitudinally polarized electron beam 
(83$\%$ polarization) from the Continuous Electron Beam Accelerator Facility 
(CEBAF) at an energy of 6.064 GeV. The data were accumulated over 40 days using 
a 6-atm-pressure, 292-mm-long, and 6-mm-diameter gaseous $^4$He target centered 
64~cm upstream of the CEBAF Large Acceptance Spectrometer (CLAS) coordinate 
center. For DVCS experiments, the CLAS baseline design \cite{Mecking:2003zu} 
was supplemented with an inner calorimeter (IC) and a solenoid magnet. The IC 
extended the photon detection acceptance of CLAS down to a polar angle of 
4$^{\circ}$. The 5-Tesla solenoid magnet in the center of which the target was 
located prevented the high-rate low-energy M{\o}ller electrons from reaching 
the CLAS drift chambers by guiding these electrons inside a tungsten shield 
placed around the beamline. 

Incoherent DVCS events were selected by requiring an electron, a proton, and at 
least one photon in the final state using the standard particle identification 
framework of the CLAS event reconstruction (see \cite{Hattawy:thesis} for 
additional details on the particle identification). Note that even though the 
DVCS reaction has only one real photon in the final state, events with more 
than one photon were not discarded at this stage. These extra photons were 
mostly soft photons from accidental coincidence which, as will be discussed 
below, the DVCS exclusivity cuts easily eliminated.  In the following 
stage, the most energetic photon was considered as the DVCS photon candidate.

Further requirements were applied to clean the identified initial set of 
incoherent DVCS events from accidental and physics background events. 
First, events were selected with $Q^2$ greater than $1$ GeV$^2$ and the 
$\gamma^*p$ invariant mass ($W=\sqrt{(q+p)^2}$, assuming that the initial 
nucleon is at rest) greater than $2$ GeV. This is a commonly accepted region of 
kinematics used by the previous DVCS experiments 
and avoids the nucleon resonance region. The squared 
transferred momentum to the recoil proton $t$, calculated from the 
four-momentum vectors of the incoming and outgoing photons, was required to be 
greater than a minimum kinematically allowed value ($t_{min}$) at given $Q^2$ 
and $W$ defined as: \begin{equation}
t_{min} 
   =-Q^{2}~\frac{2(1-~x_{B})(1~-~\sqrt{1~+~\epsilon^{2}})~+~\epsilon~^{2}}{4x_{B}(1-x_{B})+\epsilon^{2}},
\end{equation}
where $\epsilon^{2}=~\frac{4M^{2}_{p}x^{2}_{B}}{Q^{2}}$ and $M_{p}$ is the 
proton mass. This cut was applied to avoid accepting events that appear in  
unphysical regions of kinematics due to detector resolution and radiative 
effects.  We specifically use the kinematics of the photons to determine $t$ 
because the initial proton kinematics is unknown due to Fermi motion. 

\begin{figure}[tb]
\includegraphics[width=8.9cm,height=7.8cm]{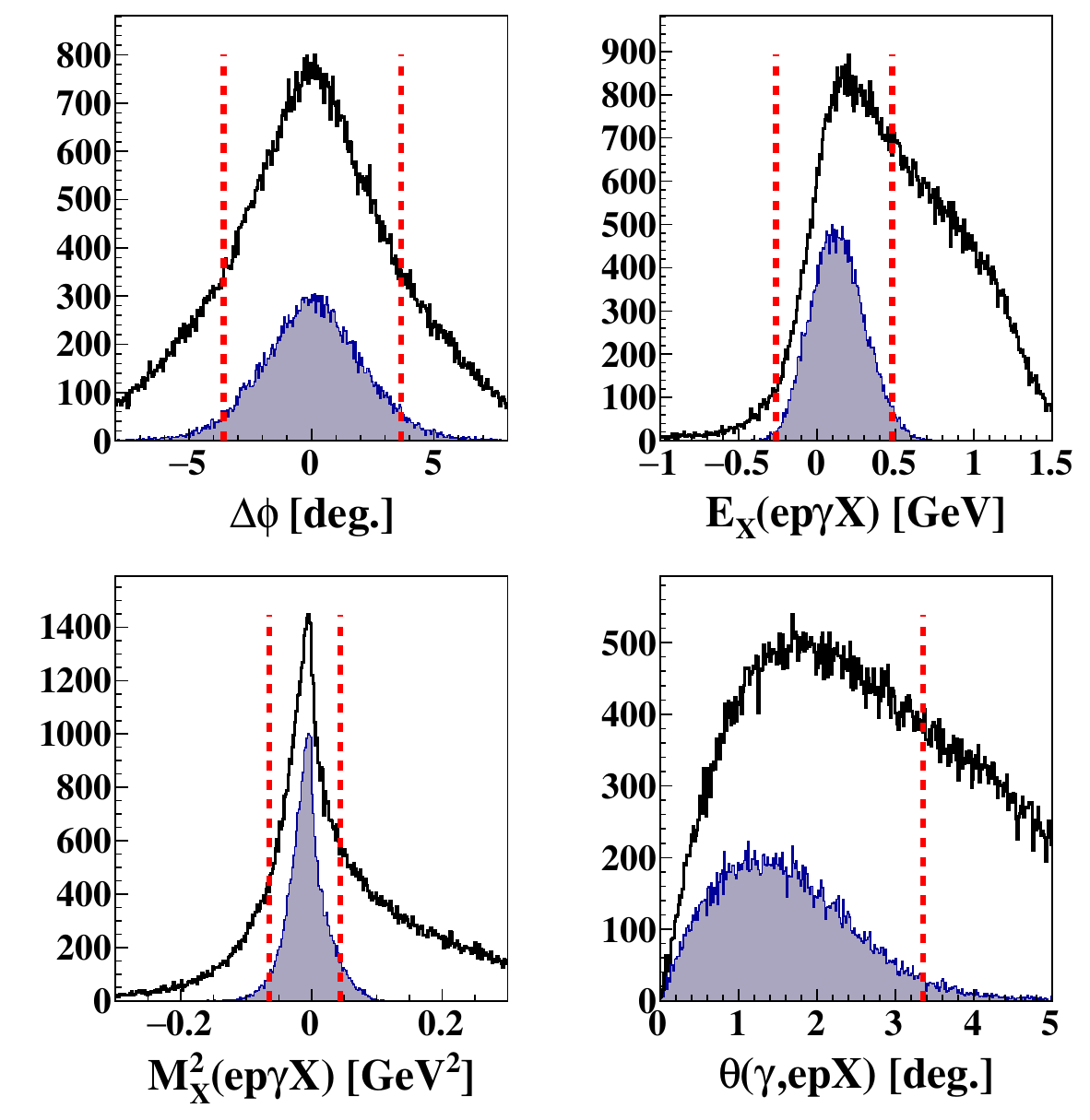}
\caption{The distributions from left to right and from top to bottom are: 
   $\Delta \phi$, missing energy, missing mass squared and the cone angle 
   ($\theta$) between the measured and the calculated photons in the $e'p'$ 
   final-state system. The incoherent DVCS exclusivity cuts are represented by 
   the vertical red-dashed lines. The black distributions represent the 
   incoherent DVCS event candidates before the exclusivity cuts. The shaded 
   distributions represent the incoherent DVCS events that passed all of these 
   cuts except the quantity plotted.}
\label{fig:kin-cuts}
\end{figure}

In the final sample, the exclusivity of the incoherent DVCS events was ensured 
by imposing a series of constraints based on the four-momentum conservation in 
the reaction $ep\rightarrow e'p'\gamma$.  These kinematical variables are: the 
coplanarity angle $\Delta\phi$ between the ($\gamma,\gamma^*$) and 
($\gamma^*$,$p'$) planes, the missing energy, mass, and transverse momentum of 
the $e'\gamma$ and $e'p'\gamma$ systems, the missing mass squared of the $e'p'$ 
system, and the angle $\theta$ between the measured photon and the missing 
momentum of the $e'p'$ system. The experimental distributions for the most relevant 
exclusivity variables are shown in Fig.~\ref{fig:kin-cuts}.  Because of the 
Fermi motion of the nucleons in the helium nucleus, the cuts indicated by the dashed lines
are slightly wider than those previously used for free proton experiments~%
\cite{Girod:2007aa}. After the corrections discussed below, the asymmetries 
appear to be stable as a function of cut width and we saw no sizable effect 
that could be related to the initial momentum of the nucleons. We also rejected 
events where a $\pi^0$ was identified by the invariant mass of two photons. At the end of this selection 
process, about 30k events passed all the requirements. 

\begin{figure*}[t!]
\includegraphics[width=13cm]{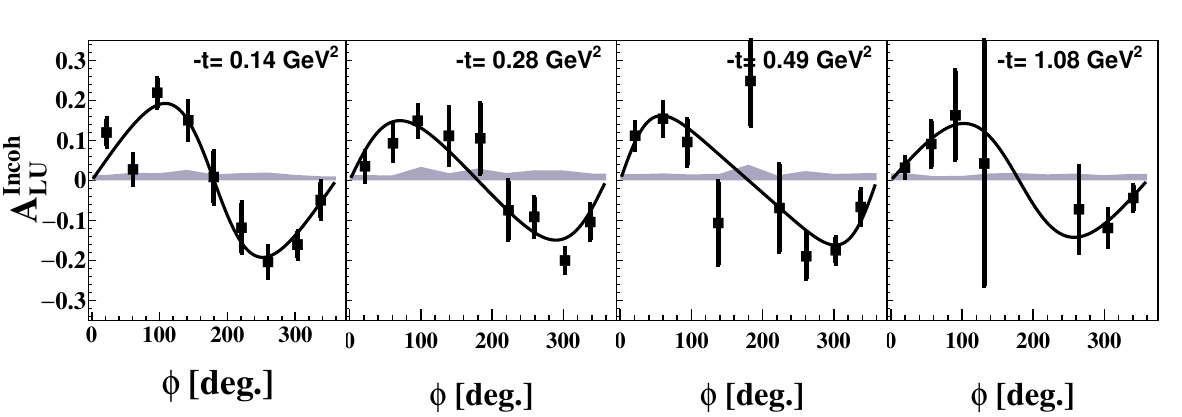}
\caption{The incoherent $A_{LU}$ as a function of $\phi$ for different $t$ 
   bins. The error bars represent the statistical uncertainties. The gray 
   bands represent the systematic uncertainties, including the normalization 
   uncertainties. The black curves are the results of our fits with the form 
   $\frac{a_{0}~\sin(\phi)}{1+ a_{1}~\cos(\phi)}$.}
\label{fig:alu}
\end{figure*}

The two main backgrounds that contributed to the event sample after the 
exclusivity cuts are due to accidental coincidences and exclusive $\pi^0$ 
production where one of the photons from the $\pi^0$ decay escapes detection.  
The contribution from accidental events, i.e., $e'p'\gamma$ collections with 
particles originating from different electron scatterings, was evaluated to be 
6.5\% by selecting events passing all our selection cuts but originating from 
different vertices. The $\pi^0$ contamination was estimated and subtracted using detector 
simulation and experimental data. From simulation, we calculated the ratio ($R 
= N^{1\gamma}_{sim}/N^{2\gamma}_{sim}$) of the number of $\pi^0$ events that 
were wrongly identified as exclusive $ep\rightarrow e'p'\gamma$ events 
($N^{1\gamma}_{sim}$) to the number of events correctly identified as exclusive 
$ep\rightarrow e'p'\pi^0$ ($N^{2\gamma}_{sim}$).  Then in each kinematical bin 
and for each beam-helicity state, the $\pi^0$-subtracted experimental DVCS 
events were calculated as $N = N^{ep\rightarrow e'p'\gamma}_{exp}- 
R~N^{ep\rightarrow e'p'\pi^0}_{exp}$, where $N^{ep\rightarrow 
e'p'\gamma}_{exp}$ ($N^{ep\rightarrow e'p'\pi^0}_{exp}$) is the number of the 
experimentally identified $ep\rightarrow e'p'\gamma$ ($ep\rightarrow 
e'p'\pi^0$) events. Depending on the kinematics, we subtracted
between 8 and 10\% of the data due to the $\pi^0$ contamination.


Experimentally, $A_{LU}$ is defined as
\begin{equation}
   A_{LU} = \frac{1}{P_{B}(1-C)} \frac{N^{+} - N^{-}}{N^{+} + N^{-} },
\end{equation}
where $N^{+}$ and $N^{-}$ are the number of DVCS events for the positive and 
negative beam-helicity states, $P_{B}$ is the longitudinal beam polarization, 
and C stands for the contamination percentage of the accidental coincidences.  

In the kinematical phase-space of our experiment, the $\phi$ dependence of 
$A_{LU}$ is most sensitive to the imaginary part of the CFFs through the $a_0$ 
term of Eq.~\ref{eq:alu-simp}, as confirmed by high statistics measurements on 
the free proton \cite{Girod:2007aa,Jo:2015ema}. In the determination of $a_0$ 
in Eq.~\ref{a0_cff}, the CFF $\mathcal{E}$ and $\mathcal{\tilde {H}}$ are 
suppressed due to form-factors and the smallness of the coefficients.  
Therefore, the dominant contribution to the BSA comes from the CFF 
$\mathcal{H}$ and hence the GPD $H$.

Due to limited statistics, the data were binned two-dimensionally into 36 bins.  
That is, four bins in one of the kinematical variable of interest ($Q^{2}$, 
$x_{B}$, or $t$) and then nine bins in the azimuthal angle ($\phi$).   
Fig.~\ref{fig:alu} presents the measured incoherent $A_{LU}$ as a function of 
$\phi$ in bins of $t$ (integrated over the full $Q^{2}$ and $x_{B}$ ranges).  
The curves on the plots are fits of the form $\frac{a_{0}~\sin(\phi)}{1+ 
a_{1}~\cos(\phi)}$. The main contributions to systematic uncertainties on these fits are from the choice of 
the DVCS exclusivity cuts (6\%) and the large bin size (7\%). The systematic 
uncertainties sum up to less than 10\% for all data points and thus always 
remain significantly smaller than the statistical uncertainities.

\begin{figure*}[t!]
\includegraphics[height=5.2cm,trim={0 0 0cm 
   0},clip]{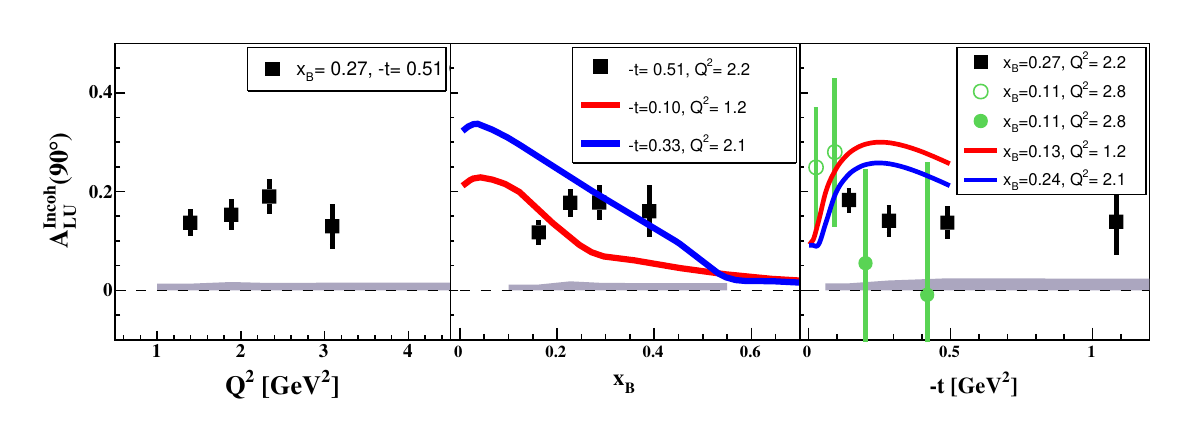}
\caption{The $Q^{2}$ (left), $x_{B}$ (middle), and $t$ dependencies (right) of
   the fitted $A_{LU}$ at $\phi$~=~90$^{\circ}$ (black squares). The error bars 
   represent the statistical uncertainties, while the gray bands represent the 
   systematic uncertainties. On the middle plot the curves are theoretical 
   calculations from \cite{simonetta_2}. On the right plot the solid (empty) 
   green circles are the HERMES $-A_{LU}$ (a positron beam was used) inclusive 
   measurements for the incoherent (coherent) enriched region 
   \cite{Airapetian:2009cga}; the curves represent theoretical calculations 
   from \cite{simonetta_2}.}
\label{fig:alu90}
\end{figure*}

Fig.~\ref{fig:alu90} presents the dependence of the fitted $A_{LU}$ values at 
$\phi$~=~90$^{\circ}$ ($a_{0}$ parameter from the individual fits in 
Fig.~\ref{fig:alu}) on the kinematical variables $Q^2$, $x_{B}$, and $t$.  
Within the given uncertainties, $A_{LU}$ does not show a strong dependence on 
$Q^2$.  The $x_{B}$ and $t$ dependencies are compared to the theoretical 
calculations performed by S.~Liuti and K.~Taneja \cite{simonetta_2}. Their 
model uses a nuclear spectral function and considers mainly the effect of the 
nucleon off-shellness. The calculations are carried out at slightly different 
kinematics than our data but still provide important guidance. The experimental 
results appear to have smaller asymmetries especially at small $x_{B}$ than the 
calculations.  These differences may arise from nuclear effects that are not 
taken into account in the model, such as long-range interactions and final 
state interactions of the knocked-out proton. On the graph for the $-t$ 
dependence, we show previous measurements by HERMES collaboration 
\cite{Airapetian:2009cga}, in which only electrons and photons were measured.  
Due to the large experimental uncertainties of the HERMES points, the two 
measurements are completely compatible.


\begin{figure}[tb]
\centering
\includegraphics[width=7.8cm]{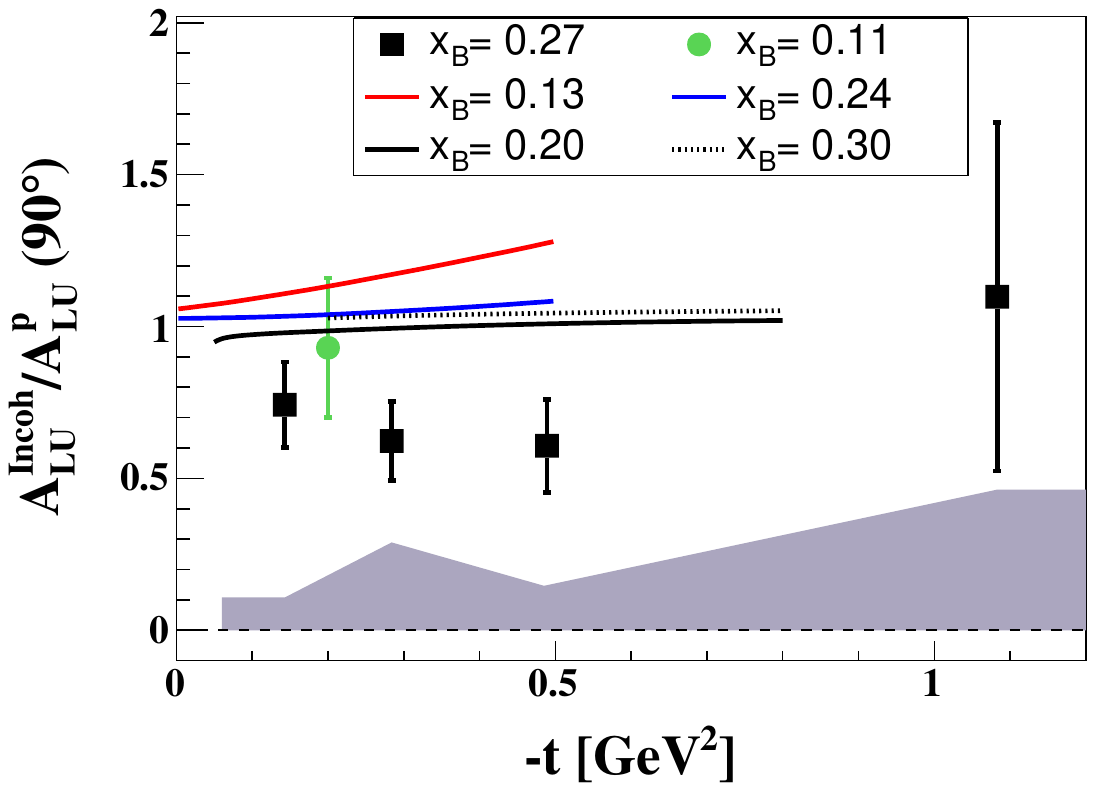}
\caption{ The $A_{LU}$ ratio of the bound to the free proton at 
   $\phi$~=~90$^{\circ}$ as a function of $t$. The black squares are from this 
   work, the green circle is the HERMES measurement \cite{Airapetian:2009cga}.  
   The error bars represent the statistical uncertainties, while the gray band 
   represents the systematic uncertainties. The blue and red curves are results 
   of off-shell calculations \cite{simonetta_2}. The solid and dashed black 
   curves are from on-shell calculations \cite{Guzey:2008fe}.} 
   \label{fig:incoh_EMC_ratio_ALU_proton}
\end{figure}

One can use the nuclear DVCS to measure a ``generalized'' EMC effect in order 
to see if significant nuclear effects are also visible within the GPD 
framework. To explore this idea, we constructed the ratio of $A_{LU}$ for bound 
protons to that on a free proton target.  
Fig.~\ref{fig:incoh_EMC_ratio_ALU_proton} presents the BSA ratio based on 
interpolation of the free proton asymmetries from CLAS \cite{Girod:2007aa} as a 
function of the kinematical variable $t$. The $A_{LU}$ ratios show 25\%-40\% 
lower asymmetries that are independent of $t$ for a bound proton compared to 
the free proton. The measurements disagree with the off-shell 
\cite{simonetta_2} and the on-shell calculations that use the medium-modified 
GPDs as calculated from the quark-meson coupling model \cite{Guzey:2008fe}. Our 
results show that an important nuclear effect is missing from the existing 
models in order to explain this strong quenching of the BSA. More theoretical 
developments will be needed to identify the origin of this quenching, in 
particular it will be important to differentiate initial from final state 
effects and how they affect the DVCS asymmetries.

In summary, we have presented the first BSA measurement associated with bound 
proton DVCS off $^4$He using an upgraded setup of the CLAS spectrometer at 
Jefferson Lab. Our results are compared to model calculations based on 
different assumptions of the nuclear medium effects at the partonic level. The 
bound-proton BSA is largely suppressed compared to the free proton BSA. This 
result is a first step in using a novel experimental method of understanding 
the properties of bound nucleons directly from the basic degrees of freedom of 
QCD, quarks and gluons. Planned experiments at Jefferson Lab will continue 
and extend these studies of the bound nucleon structure using DVCS. We have an 
experimental program called ALERT using the CLAS12 detector in the Hall-B of 
Jefferson Lab. These experiments will improve the DVCS measurements with the 
detection of nuclear fragments to better control the final state interactions 
and the initial state kinematics of the bound nucleon.


The authors acknowledge the staff of the Accelerator and Physics Divisions at 
the Thomas Jefferson National Accelerator Facility who made this experiment 
possible. This work was supported in part by the Chilean Comisi\'on Nacional de 
Investigaci\'on Cient\'ifica y Tecnol\'ogica (CONICYT), by CONICYT PIA grant 
ACT1413, the Italian Instituto Nazionale di Fisica Nucleare, the French Centre 
National de la Recherche Scientifique, the French Commissariat \`a l'Energie 
Atomique, the U.S.  Department of Energy under Contract No. DE-AC02-06CH11357, 
the United Kingdom Science and Technology Facilities Council (STFC), the 
Scottish Universities Physics Alliance (SUPA), the National Research Foundation 
of Korea, and the Office of Research and Economic Development at Mississippi 
State University.  M.~Hattawy also acknowledges the support of the Consulat 
G\'en\'eral de France \`a J\'erusalem.  The Southeastern Universities Research 
Association operates the Thomas Jefferson National Accelerator Facility for the 
United States Department of Energy under Contract No. DE-AC05-06OR23177.

\end{document}

%% file: author_list.tex
\newcommand*{\ANL}{Argonne National Laboratory, Argonne, Illinois 60439}
\newcommand*{\ANLindex}{1}
\affiliation{\ANL}
\newcommand*{\ASU}{Arizona State University, Tempe, Arizona 85287-1504}
\newcommand*{\ASUindex}{2}
\affiliation{\ASU}
\newcommand*{\CANISIUS}{Canisius College, Buffalo, NY}
\newcommand*{\CANISIUSindex}{3}
\affiliation{\CANISIUS}
\newcommand*{\CMU}{Carnegie Mellon University, Pittsburgh, Pennsylvania 15213}
\newcommand*{\CMUindex}{4}
\affiliation{\CMU}
\newcommand*{\CUA}{Catholic University of America, Washington, D.C. 20064}
\newcommand*{\CUAindex}{5}
\affiliation{\CUA}
\newcommand*{\SACLAY}{IRFU, CEA, Universit\'e Paris-Saclay, F-91191 
Gif-sur-Yvette, France}
\newcommand*{\SACLAYindex}{6}
\affiliation{\SACLAY}
\newcommand*{\CNU}{Christopher Newport University, Newport News, Virginia 23606}
\newcommand*{\CNUindex}{7}
\affiliation{\CNU}
\newcommand*{\UCONN}{University of Connecticut, Storrs, Connecticut 06269}
\newcommand*{\UCONNindex}{8}
\affiliation{\UCONN}
\newcommand*{\DUKE}{Duke University, Durham, North Carolina 27708-0305}
\newcommand*{\DUKEindex}{9}
\affiliation{\DUKE}
\newcommand*{\FU}{Fairfield University, Fairfield CT 06824}
\newcommand*{\FUindex}{10}
\affiliation{\FU}
\newcommand*{\FERRARAU}{Universita' di Ferrara , 44121 Ferrara, Italy}
\newcommand*{\FERRARAUindex}{11}
\affiliation{\FERRARAU}
\newcommand*{\FIU}{Florida International University, Miami, Florida 33199}
\newcommand*{\FIUindex}{12}
\affiliation{\FIU}
\newcommand*{\FSU}{Florida State University, Tallahassee, Florida 32306}
\newcommand*{\FSUindex}{13}
\affiliation{\FSU}
\newcommand*{\Genova}{Universit$\grave{a}$ di Genova, 16146 Genova, Italy}
\newcommand*{\Genovaindex}{14}
\affiliation{\Genova}
\newcommand*{\GWUI}{The George Washington University, Washington, DC 20052}
\newcommand*{\GWUIindex}{15}
\affiliation{\GWUI}
\newcommand*{\ISU}{Idaho State University, Pocatello, Idaho 83209}
\newcommand*{\ISUindex}{16}
\affiliation{\ISU}
\newcommand*{\INFNFE}{INFN, Sezione di Ferrara, 44100 Ferrara, Italy}
\newcommand*{\INFNFEindex}{17}
\affiliation{\INFNFE}
\newcommand*{\INFNFR}{INFN, Laboratori Nazionali di Frascati, 00044 Frascati, Italy}
\newcommand*{\INFNFRindex}{18}
\affiliation{\INFNFR}
\newcommand*{\INFNGE}{INFN, Sezione di Genova, 16146 Genova, Italy}
\newcommand*{\INFNGEindex}{19}
\affiliation{\INFNGE}
\newcommand*{\INFNRO}{INFN, Sezione di Roma Tor Vergata, 00133 Rome, Italy}
\newcommand*{\INFNROindex}{20}
\affiliation{\INFNRO}
\newcommand*{\INFNTUR}{INFN, Sezione di Torino, 10125 Torino, Italy}
\newcommand*{\INFNTURindex}{21}
\affiliation{\INFNTUR}
\newcommand*{\ORSAY}{Institut de Physique Nucl\'eaire, IN2P3-CNRS, Universit\'e 
Paris-Sud, Universit\'e Paris-Saclay, F-91406 Orsay, France}
\newcommand*{\ORSAYindex}{22}
\affiliation{\ORSAY}
\newcommand*{\ITEP}{Institute of Theoretical and Experimental Physics, Moscow, 117259, Russia}
\newcommand*{\ITEPindex}{23}
\affiliation{\ITEP}
\newcommand*{\JMU}{James Madison University, Harrisonburg, Virginia 22807}
\newcommand*{\JMUindex}{24}
\affiliation{\JMU}
\newcommand*{\KNU}{Kyungpook National University, Daegu 41566, Republic of Korea}
\newcommand*{\KNUindex}{25}
\affiliation{\KNU}
\newcommand*{\LAMAR}{Lamar University, 4400 MLK Blvd, PO Box 10009, Beaumont, Texas 77710}
\newcommand*{\LAMARindex}{26}
\affiliation{\LAMAR}
\newcommand*{\LPSC}{LPSC, Universit\'e Grenoble-Alpes, CNRS/IN2P3, 38026 
Grenoble, France}
\newcommand*{\LPSCindex}{27}
\affiliation{\LPSC}
\newcommand*{\MISS}{Mississippi State University, Mississippi State, MS 39762-5167}
\newcommand*{\MISSindex}{28}
\affiliation{\MISS}
\newcommand*{\UNH}{University of New Hampshire, Durham, New Hampshire 03824-3568}
\newcommand*{\UNHindex}{29}
\affiliation{\UNH}
\newcommand*{\NSU}{Norfolk State University, Norfolk, Virginia 23504}
\newcommand*{\NSUindex}{30}
\affiliation{\NSU}
\newcommand*{\OHIOU}{Ohio University, Athens, Ohio  45701}
\newcommand*{\OHIOUindex}{31}
\affiliation{\OHIOU}
\newcommand*{\ODU}{Old Dominion University, Norfolk, Virginia 23529}
\newcommand*{\ODUindex}{32}
\affiliation{\ODU}
\newcommand*{\URICH}{University of Richmond, Richmond, Virginia 23173}
\newcommand*{\URICHindex}{33}
\affiliation{\URICH}
\newcommand*{\ROMAII}{Universita' di Roma Tor Vergata, 00133 Rome Italy}
\newcommand*{\ROMAIIindex}{34}
\affiliation{\ROMAII}
\newcommand*{\MSU}{Skobeltsyn Institute of Nuclear Physics, Lomonosov Moscow State University, 119234 Moscow, Russia}
\newcommand*{\MSUindex}{35}
\affiliation{\MSU}
\newcommand*{\SCAROLINA}{University of South Carolina, Columbia, South Carolina 29208}
\newcommand*{\SCAROLINAindex}{36}
\affiliation{\SCAROLINA}
\newcommand*{\TEMPLE}{Temple University,  Philadelphia, PA 19122 }
\newcommand*{\TEMPLEindex}{37}
\affiliation{\TEMPLE}
\newcommand*{\JLAB}{Thomas Jefferson National Accelerator Facility, Newport News, Virginia 23606}
\newcommand*{\JLABindex}{38}
\affiliation{\JLAB}
\newcommand*{\UTFSM}{Universidad T\'{e}cnica Federico Santa Mar\'{i}a, Casilla 110-V Valpara\'{i}so, Chile}
\newcommand*{\UTFSMindex}{39}
\affiliation{\UTFSM}
\newcommand*{\GLASGOW}{University of Glasgow, Glasgow G12 8QQ, United Kingdom}
\newcommand*{\GLASGOWindex}{40}
\affiliation{\GLASGOW}
\newcommand*{\YORK}{University of York, York YO10 5DD, United Kingdom}
\newcommand*{\YORKindex}{41}
\affiliation{\YORK}
\newcommand*{\VT}{Virginia Tech, Blacksburg, Virginia   24061-0435}
\newcommand*{\VTindex}{42}
\affiliation{\VT}
\newcommand*{\VIRGINIA}{University of Virginia, Charlottesville, Virginia 22901}
\newcommand*{\VIRGINIAindex}{43}
\affiliation{\VIRGINIA}
\newcommand*{\WM}{College of William and Mary, Williamsburg, Virginia 23187-8795}
\newcommand*{\WMindex}{45}
\affiliation{\WM}
\newcommand*{\YEREVAN}{Yerevan Physics Institute, 375036 Yerevan, Armenia}
\newcommand*{\YEREVANindex}{46}
\affiliation{\YEREVAN}

\newcommand*{\NOWGWUI}{ The George Washington University, Washington, DC 20052}
\newcommand*{\NOWSAD}{ Imam Abdulrahman Bin Faisal University, Industrial 
Jubail 31961, Saudi Arabia}
\newcommand*{\NOWISU}{ Idaho State University, Pocatello, Idaho 83209}
\newcommand*{\NOWINFNGE}{ INFN, Sezione di Genova, 16146 Genova, Italy}


\author {M.~Hattawy}
\affiliation{\ANL}
\affiliation{\ORSAY}
\affiliation{\ODU}
\author {N.A.~Baltzell}
\affiliation{\ANL}
\affiliation{\ODU}
\affiliation{\JLAB}
\author {R.~Dupr\'{e}}
\affiliation{\ANL}
\affiliation{\ORSAY}
\author {S.~B\"{u}ltmann} 
\affiliation{\ODU}
\author{R.~De~Vita} 
\affiliation{\INFNGE}
\author {A.~El~Alaoui} 
\affiliation{\ANL}
\affiliation{\UTFSM}
\author {L.~El~Fassi} 
\affiliation{\ANL}
\affiliation{\MISS}
\author{H.~Egiyan}
\affiliation{\JLAB}
\author{F.X.~Girod} 
\affiliation{\JLAB}
\author {M.~Guidal} 
\affiliation{\ORSAY}
% ...

\collaboration{The CLAS Collaboration}
\noaffiliation

%% file: incoherent_prl.bbl
\begin{thebibliography}{99}

\bibitem{Hofstadter:1955ae} R.~Hofstadter and R.~W.~McAllister,
Phys.\ Rev.\  {\bf 98}, 217 (1955).

\bibitem{Perdrisat:2006hj} C.~F.~Perdrisat, V.~Punjabi and M.~Vanderhaeghen,
Prog.\ Part.\ Nucl.\ Phys.\  {\bf 59}, 694 (2007).

\bibitem{pdg} M.~Tanabashi {\it et al.} (Particle Data Group), Phys.\ Rev.\ D 
   {\bf 98}, 030001 (2018).

\bibitem{EMC_first}
   J.~J.~Aubert {\it et al.}, 
      Phys.\ Lett.\, vol.\ B { \bf 123}, pp. 275–278 (1983).

\bibitem{Arneodo:1992wf} 
  M.~Arneodo,
  Phys.\ Rept.\  {\bf 240}, 301 (1994).

\bibitem{Geesaman:1995yd} 
  D.~F.~Geesaman, K.~Saito and A.~W.~Thomas,
  Ann.\ Rev.\ Nucl.\ Part.\ Sci.\  {\bf 45}, 337 (1995).

\bibitem{Norton:2003cb} 
  P.~R.~Norton,
  Rept.\ Prog.\ Phys.\  {\bf 66}, 1253 (2003).

\bibitem{Hen:2016kwk} 
  O.~Hen {\it et al.},
  Rev.\ Mod.\ Phys.\  {\bf 89}, no. 4, 045002 (2017).

\bibitem{Mueller:1998fv} D. Mueller, D. Robaschik, B. Geyer, F.M. Dittes, and 
   J.  Horejsi,
Fortsch.\ Phys. {\bf 42}, 101 (1994).
  
\bibitem{Ji:1996ek} 
X.D. Ji,
Phys.\ Rev.\ Lett. {\bf 78}, 610 (1997).

\bibitem{Ji:1996nm} 
X.D. Ji,
Phys.\ Rev.\ D {\bf 55}, 7114 (1997).

\bibitem{Radyushkin:1996nd}
A.V. Radyushkin,
Phys.\ Lett.\  B {\bf 380}, 417 (1996).

\bibitem{Radyushkin:1997ki} 
A.V. Radyushkin,
Phys.\ Rev.\ D {\bf 56}, 5524 (1997).

\bibitem{Stepanyan:2001sm}
S.~Stepanyan {\it et al.} [CLAS Collaboration],
Phys.\ Rev.\ Lett. {\bf 87}, 182002 (2001).

\bibitem{Airapetian:2001yk} A.~Airapetian {\it et al.} [HERMES Collaboration],
 Phys.\ Rev.\ Lett.\  {\bf 87}, 182001 (2001).

\bibitem{Airapetian:2006zr} A.~Airapetian {\it et al.} [HERMES Collaboration],
Phys.\ Rev.\ D {\bf 75}, 011103 (2007).

\bibitem{Chekanov:2003ya}
S. Chekanov {\it et al.} [ZEUS Collaboration],
Phys.\ Lett.\  B {\bf 573}, 46 (2003).

\bibitem{Aktas:2005ty}
A. Aktas {\it et al.} [H1 Collaboration],
Eur.\ Phys.\ J.\ C {\bf 44}, 1 (2005).

\bibitem{Chen:2006na} 
S.~Chen {\it et al.} [CLAS Collaboration],
Phys.\ Rev.\ Lett.\ {\bf 97}, 072002 (2006).

\bibitem{Defurne:2015kxq}
M.~Defurne {\it et al.} [Jefferson Lab Hall A Collaboration],
  Phys.\ Rev.\ C {\bf 92}, no. 5, 055202 (2015).

\bibitem{Girod:2007aa} 
F.X. Girod {\it et al.} [CLAS Collaboration],
Phys.\ Rev.\ Lett. {\bf 100}, 162002 (2008).

\bibitem{Mazouz:2007aa} 
   M.~Mazouz {\it et al.} [Jefferson Lab Hall A Collaboration],
   Phys.\ Rev.\ Lett.\  {\bf 99}, 242501 (2007).

\bibitem{Gavalian:2009} 
G. Gavalian {\it et al.} [CLAS Collaboration],
Phys.\ Rev.\ C {\bf 80}, 035206 (2009).

\bibitem{Seder:2015} 
E. Seder {\it et al.} [CLAS Collaboration],
Phys.\ Rev.\ Lett. {\bf 114}, 032001 (2015).

\bibitem{Pisano:2015} 
S.~Pisano {\it et al.} [CLAS Collaboration],
Phys.\ Rev.\ D {\bf 91}, 052014 (2015).

\bibitem{Jo:2015ema} H.~S.~Jo {\it et al.} [CLAS Collaboration],
  Phys.\ Rev.\ Lett.\  {\bf 115}, no. 21, 212003 (2015).

\bibitem{Guidal:2013rya} M.~Guidal, H.~Moutarde, and M.~Vanderhaeghen,
Rep.\ Prog.\ Phys.\  {\bf 76}, 066202 (2013).

\bibitem{Dupre:2016mai} R.~Dupr\'{e}, M.~Guidal and M.~Vanderhaeghen,
 Phys.\ Rev.\ D {\bf 95}, no. 1, 011501 (2017).

\bibitem{Dupre:2015jha} 
  R.~Dupr\'e and S.~Scopetta,
  Eur.\ Phys.\ J.\ A {\bf 52}, no. 6, 159 (2016).

\bibitem{Hattawy:2017woc} M.~Hattawy {\it et al.} [CLAS Collaboration],
   Phys.\ Rev.\ Lett.\  {\bf 119}, no. 20, 202004 (2017).


\bibitem{simonetta_2}
S.~Liuti and K.~Taneja, Phys.\ Rev.\ C {\bf 72}, 032201 (2005).

\bibitem{Guzey:2006xi} V.~Guzey and T.~Teckentrup, Phys.\ Rev.\ D {\bf 74}, 
   054027 (2006)

\bibitem{Guzey:2008fe} V.~Guzey, A.~W.~Thomas and K.~Tsushima,
  Phys.\ Lett.\ B {\bf 673}, 9 (2009).

\bibitem{JSeely}
J. Seely {\it et al.}, Phys.\ Rev.\ Lett.\ {\bf 103}, 202301 (2009).

\bibitem{Airapetian:2009cga} A.~Airapetian {\it et al.} [HERMES Collaboration],
Phys.\ Rev.\ C {\bf 81}, 035202 (2010).

\bibitem{Freund_Collins}
J.C.~Collins and A.~Freund, Phys.\ Rev.\ D {\bf 59}, 074009 (1999).

\bibitem{Ji_Osborne}
   X.-D.~Ji and J.~Osborne, Phys.\ Rev.\ D {\bf 58}, 094018 (1998).

\bibitem{Belitsky:2001ns}
A.~V.~Belitsky, D.~Mueller and A.~Kirchner, Nucl.\ Phys.\ B {\bf 629}, 323 
(2002). Updated in: A.~V.~Belitsky and D.~Mueller, Phys.\ Rev.\ D {\bf 82}, 
074010 (2010).


\bibitem{Hafidi:2008pr} K.~Hafidi {\it et al.},
   proposal PR-08-024 to JLab PAC33 
   (\url{https://www.jlab.org/exp_prog/proposals/08/PR-08-024.pdf}).

\bibitem{Mecking:2003zu} B.~A.~Mecking {\it et al.},
   Nucl.\ Instrum.\ Meth.\ A {\bf 503}, 513 (2003).

\bibitem{Hattawy:thesis}
M.~Hattawy, Ph.D. thesis, Universit{\'e} Paris Sud - Paris XI, France, 2015 
[Institution Report No. 2015PA112161].


\end{thebibliography}
